\documentclass[prb,twocolumn,showpacs]{revtex4}
\usepackage{amsmath}
\usepackage[dvips]{graphicx}
\usepackage{amssymb}
\usepackage{color}
\usepackage{epstopdf}
\DeclareGraphicsExtensions{.pdf,.eps,.png,.jpg,.mps}

\begin{document}

\title{Van der Waals interaction in magnetic bilayer graphene nanoribbons}

\author{H. Santos$^{1,2,3}$}
\email{hernan.santos@icmm.csic.es}
\altaffiliation[Present address: ]{Instituto de Ciencia de Materiales de Madrid, CSIC.}
\author{A. Ayuela$^{3}$}
\author{L. Chico$^1$}
\author{Emilio Artacho$^{2,4,5,6}$}
 
\affiliation{
$^1$ Instituto de Ciencia de Materiales de Madrid, CSIC, Cantoblanco, 28049 Madrid, Spain\\
$^2$ Department of Earth Sciences, University of Cambridge, Downing Street, Cambridge CB2 3EQ, United Kingdom\\
$^3$ Centro de F\'\i sica de Materiales CFM-CPM CSIC-UPV/EHU, Departamento de F\'\i sica de Materiales (Facultad de  Qu\'\i mica, UPV) and Donostia International Physics Center, 20080 San Sebasti\'an/Donostia, Spain\\
$^4$ Nanogune and DIPC, Tolosa Hiribidea 76, 20018 San Sebasti\'an, Spain\\
$^5$ Basque Foundation for Science, Ikerbasque, 48011 Bilbao, Spain\\
$^6$ Cavendish Laboratory, University of Cambridge, Cambridge CB3 0HE, United Kingdom}

\date{\today}

\begin{abstract}

We study  the interaction energy  between two graphene  nanoribbons by first principles 
calculations, 
including van der Waals interactions and spin  polarization.  For  ultranarrow zigzag  nanoribbons,  the direct stacking  is even  more stable  than the Bernal stacking,  competing in  energy for 
wider ribbons. 
This behavior  is  due  to  the magnetic  interaction between edge states.  We relate  the reduction of the magnetization in zigzag  nanoribbons with  increasing  ribbon width  to the  structural changes  produced by  the  magnetic interaction,  and we show that  when deposited on  a substrate, zigzag bilayer ribbons  remain magnetic for larger widths.

\end{abstract}
\maketitle

\section{Introduction}

Charge carriers in graphene follow a linear dispersion relation close to the Fermi energy. For this reason, they are considered as massless fermions obeying Dirac's equation. \cite{transport} 
When
several layers of graphene are piled up together, their electronic and transport properties can be dramatically modified, depending on the stacking arrangement and the number of layers.\cite{Heinz} There  
are
several
possible stacking arrangements in bilayer 
graphene, 
the most symmetric cases being direct (AA) 
 and Bernal (AB) stackings. Most theoretical studies have focused on the AB stacking, because it is that of graphite, being the lowest energy configuration for the three-dimensional crystal.  \cite{charlier1} However, 
the
AA stacking has been 
observed in experiments on few-layer graphene, and it should also be 
considered 
in bilayer stackings.\cite{Norimatsu,jae_kap,Liu,Ohto} For example, 
the
AA and AB stackings have been observed indistinctly at 
the
graphene edges in samples grown on SiC. \cite{Norimatsu} 
In fact,
bilayer graphene with AA stacking has also been synthesized, and observed by transmission electron microscopy (TEM).  \cite{jae_kap,Liu} Due to the differences in the electronic properties of bilayer AA and AB, the change 
between stackings by a relative displacement of the layers has even been proposed as the key mechanism for a switch device.\cite{Ohto,Jhon1,Jhon2}

Graphene nanoribbons are graphene strips of nanometric width and arbitrary length, with electronic properties depending on their edges and widths.\cite{son_nature,Brey_2006} They are considered as potential materials for future nanoelectronics because they can behave as metals or semiconductors, making possible the design of electronic elements based solely on them.  The simplest nanoribbon geometries are those with zigzag and armchair edges, which have been studied extensively.\cite{Brey_2006} Other edges terminations are possible, but they can be mapped onto three basic types, the armchair being the only one without edge states.\cite{Akhmerov,Jaskolski2011} These localized states close to the Fermi energy are responsible for the magnetic and transport properties of zigzag graphene ribbons, and they are the origin of defect-related interface bands in graphene junctions.\cite{Santos2009} Within a simple tight-binding model, armchair graphene nanoribbons (AGNRs) can be either metallic or semiconducting depending on their width, \cite{nakada} whereas zigzag graphene nanoribbons (ZGNRs) are metallic with edge states. \cite{son_2006} More realistic calculations yield all semiconducting armchair ribbons. \cite{Ezawa,Barone} 
 With regard to ZGNRs, the inclusion of electronic interactions reveals a ferromagnetic order of the magnetic moments at each edge, with an edge-edge antiferromagnetic coupling that 
opens
a small gap. In fact, this 
magnetic characteristic
makes ZGNRs interesting for spintronic devices. \cite{alicante,santos}

In bilayer graphene nanoribbons ({\it b}-GNR) both, edges and stacking order, determine the electronic and magnetic properties. Even though in few-layer samples there are multiple possibilities for the stacking arrangements, the majority of previous theoretical studies have focused on the AB stacking.\cite{McCann,McCann2,Nilsson,banerjee,fazzio}
The interaction between the edges of zigzag bilayer graphene ribbons determines the survival of magnetism. A combined first-principles and tight-binding approach was used to study the electronic properties in armchair and zigzag GNR.\cite{banerjee} Because these authors find  an important dependence on the functional employed, they fix their distance to graphite for zigzag GNR.  Their magnetism is thus masked, as its survival depends on the layer-layer distance. An attempt to relax the edges was considered using a local spin density  (LSD) approach within density functional theory.\cite{Kim} 
They found that, for wide bilayer zigzag nanoribbons, the total magnetic moment is zero. 

Previous works do not consider van der Waals (vdW) forces. In order to relax bilayer ribbons, an explicit description of the vdW interaction must be included beyond LSD.
 When these long-range interactions  
 are included, the electronic densities between the layers are rearranged, and this yields variations on the interlayer distances.  Such vdW interaction is included at a simple level in Ref. \onlinecite{fazzio} and the edge magnetism disappears for small ribbon widths. However, we should note that vdW interaction is included in an effective way, modifying the atomic potentials. Other implementations using a fully nonlocal van der Waals density functional must thus be checked.

In this work we study the properties of bilayer zigzag ribbons where all the edge carbon atoms are passivated by hydrogen. We include 
 van der Waals dispersion forces with the fully non-local density functional recently proposed from first-principles,\cite{LMKLL} within the family of functionals based on Ref. \onlinecite{DRSLL}, as recently factorized for efficiency. \cite{Roman&Soler} 
In Section II we describe the computational details.
We investigate the stability of Bernal and direct stackings in {\it b}-GNR, focusing on the magnetic interaction between edges and on the interplay between magnetism and structural changes in narrow zigzag ribbons.
%
Section III describes the systems studied, and shows our results, presenting the binding energies, and magnetic and structural changes in zigzag bilayer nanoribbons. 
Our main conclusion is that direct stacking competes with Bernal stacking below a critical ribbon width, and we show that the magnetic coupling between edge states in the different ribbons plays a key role in such competition. 
 Indeed, for ultranarrow ribbons, the direct stacking has the lowest total energy and largest binding energy. Furthermore, the structural distorsion at the edges due to this interaction makes the magnetization negligible in bilayer ribbons, causing metallization. However, when deposited on a substrate, the structural deformation is reduced, thus maintaining the edge magnetism for larger ribbon widths.
 We finish with a brief summary in Section IV.

\section{Computational details}
\label{sec:comp}

First principles calculations are performed using the SIESTA code with spin polarization.\cite{siesta1} We use the van der Waals functional parametrized by Lee {\it  et al.} 
 (vdW-DF2),\cite{LMKLL} which is a second version of the original vdW-DF functional by Dion and coworkers.\cite{DRSLL}
   The factorization proposed in Ref. \onlinecite{Roman&Soler} represents a very substantial efficiency improvement in the evaluation of the exchange-correlation potential and energy, thus enabling first-principles van der Waals calculations for any system accessible to usual generalized gradient approximations GGAs. 
 We check that the interlayer space in graphite is in agreement with previous calculations.\cite{LMKLL} The results presented below have been performed using the functional vdW-DF2, but we have checked that other functionals implemented in the SIESTA code preserve the main features found employing vdW-DF2. Our choice of functional is motivated by the fact that vdW-DF2 gives more realistic binding energies for bilayer graphene when compared to experimental works. The electron-ion interactions use norm-conserving nonlocal Troullier-Martins pseudopotentials\cite{Troullier} generated with the atomic configuration [He]$2s^2$$2p^2$ taken as reference  with a radius cutoff of 1.25 \AA \  for $s$, $p$, $d$ and $f$ orbitals. 
Spin polarized calculations 
normally
require a fine sampling of the Brillouin zone, which we performed with a Monkhorst-Pack scheme of $30\times1\times1$ $k$-points.\cite{Monkhorst} 
The real-space grid for matrix-element computations \cite{siesta1} uses 
 an energy cutoff of 350 Ry. The structure was relaxed by conjugate gradients optimization until forces were smaller than 0.01 eV/\AA.
  Periodic boundary conditions were applied, so we use large enough supercell parameters (15 \AA) in the directions perpendicular to the ribbon's long axis to avoid spurious interactions between adjacent ribbons.  
All the carbon atoms at the edges are passivated by hydrogen. %

\section{Results}
\label{sec:rec}

\subsection{Ingredients: monolayer zigzag nanoribbons}

Before undertaking the calculation of bilayer nanoribbons, we have first verified that our approach gives reasonable results for monolayer zigzag nanoribbons. For these, the key parameters to determine the electronic behavior at the Fermi energy are both the edge shape and the ribbon width. We have used two initial magnetic configurations for the edges of ZGNR: either ferromagnetic ({\it fm}), with aligned spin polarizations, or antiferromagnetic ({\it afm}), with antiparallel spin polarizations. 
Notice that all atoms in the same edge are ferromagnetic coupled. \cite{son_2006}
 We have performed calculations of ZGNRs with several widths and both {\it afm} or {\it fm} orderings. We found that the {\it afm} order is always more stable than the {\it fm}, and that their energy difference decreases with increasing ribbon width, as in previous calculations.\cite{son_2006}

\begin{figure}[h!] 
   \centering
     \includegraphics[width=8.4cm]{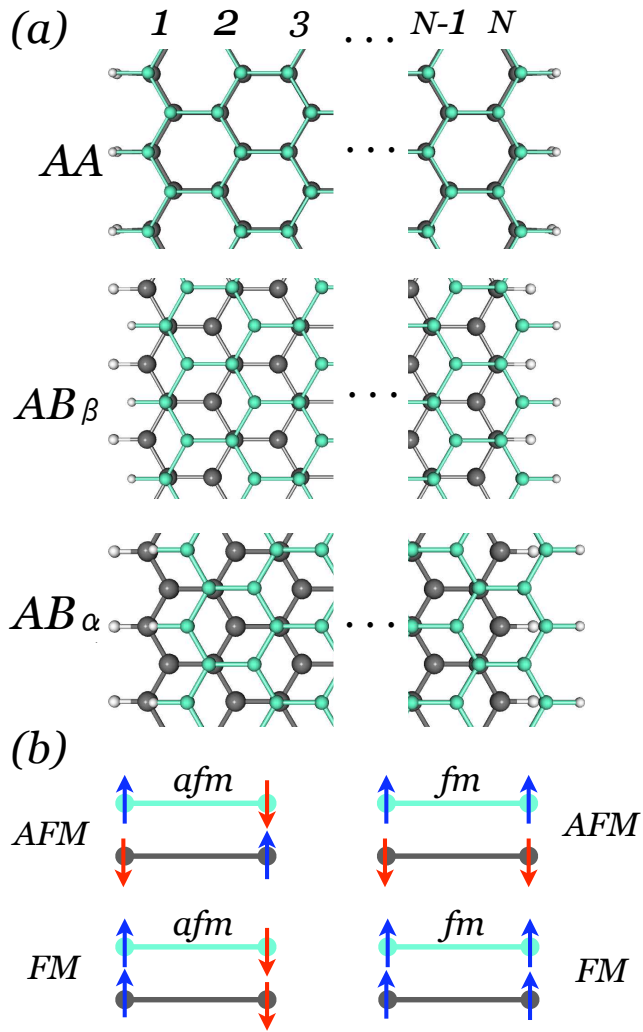}
     \caption{(Color online) (a) Schematic packing of zigzag bilayer graphene nanoribbons: AA (above), AB$_\beta$ (middle), and AB$_\alpha$(below). The dark gray layer corresponds to the bottom layer and the light gray (blue) layer correspond to the upper layer.  Hydrogen atoms are denoted by white balls. 
The ribbon width $N$ is given by the number of dimers (zigzag chains) from edge to edge. (b) Magnetic configurations of the edge states in {\it b}-ZGNR. Interlayer (intralayer) coupling can be 
either
ferromagnetic [FM({\it fm})] or antiferromagnetic [AFM({\it afm})].
Notice that one single edge is always ferromagnetic, i.e., all the spins along the same edge are parallel.
}
   \label{fig1}
\end{figure}
\subsection{Bilayer zigzag nanoribbons}

\subsubsection{Binding energies and stable configurations}

As in infinite bilayer graphene, we have to 
look at
different stacking orders for bilayer graphene nanoribbons ({\it b}-ZGNRs).  Fig. \ref{fig1} (a) shows the three stackings
 considered in this work. The top panel of the figure depicts an example of direct (AA) stacking. 
Two 
types of AB stacking have to be considered, according to the relative position of their edges. The medium and bottom panels of  Fig.\ref{fig1} (a) show the so-called AB$_\beta$ and   AB$_\alpha$ stackings for zigzag ribbons. We identify the ribbon width by $N$, being the number of zigzag chains from edge to edge. \cite{nakada} 

As the edges of {\it b}-ZGNRs have magnetization, we have to study both the intralayer and interlayer (i.e., layer-to-layer) magnetic couplings in these ribbons.  We study four possible magnetic configurations for all the stackings considered, as depicted in Fig. \ref{fig1} (b).  In order to distinguish in our notation between intralayer and interlayer coupling, we use capital letters for the layer-to-layer coupling, and lower-case letters to label the intralayer coupling.  The AFM-{\it afm} configuration (upper left diagram of Fig. \ref{fig1} (b)) thus has both antiferromagnetic intralayer and interlayer coupling. The bottom-left diagram, the FM-{\it afm},  shows two {\it afm}-coupled GNR layers with FM interlayer coupling. The third and fourth configurations, AFM-{\it fm} and FM-{\it fm}, shown in the top and bottom right diagrams of Fig.  \ref{fig1} (b), have both {\it fm} intralayer coupling with AFM or FM interlayer coupling respectively.

We start from one of the four initial spin configurations for {\it b}-ZGNRs described above
in a range of distances,
 and then we relax the 
minimum
geometry so that the system evolves in principle towards similar magnetic configurations with lower energy. The converged solutions for different initial guesses are very close in 
energy. Each converged magnetic configuration can be viewed as a possible metastable solution.\cite{Marom} It
is likely that external conditions, such as magnetic and electric fields, can stabilize the system into in a configuration different from the energy minimum. 

However, for the AA stacking, the FM-({\it afm}, {\it fm}) magnetic initial guesses do not yield a stable solution: as the layers 
become
closer, the electron density flips during self-consistency to the AFM ground state. The same happens for the AB stacking,  where the FM-{\it afm} cases flip to AFM-{\it afm} solutions. Since for the AB stackings the atoms of an edge are not exactly on top of the atoms of the other, we obtained a large number of metastable magnetic alignments.  

\begin{figure}[htbp] 
  \centering
    \includegraphics[clip,width=7cm,angle=-90]{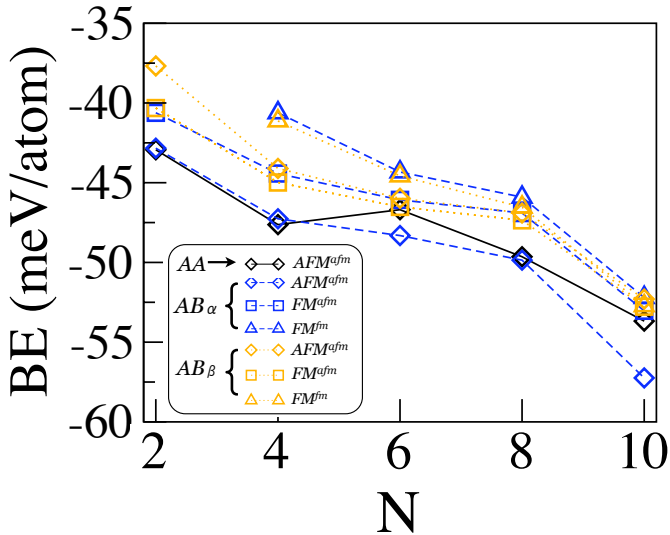}
  \caption{(Color online) Binding energy (BE) of {\it b}-ZGNR as a function of the ribbon width $N$  after full relaxation. It is noteworthy that the AB$_\alpha$ and AA stackings have the largest absolute values of the binding energies.
    }
  \label{fig2}
 \end{figure}

From the total energies, we calculate the binding energy (BE) as the difference between the energy of the coupled bilayer and the two isolated monolayers in the most stable configuration, i.e., the antiferromagnetic ({\it afm}) alignment.\cite{BE}
  The binding energy is related to the strength of the layer-layer interaction; it is given in meV per atom in a layer.\cite{Grimme,Podeszwa} Figure \ref{fig2} shows the binding energy of {\it b}-ZGNRs with widths $N$ ranging from 2 to 10, for several magnetic configurations between edges.
The binding energies that we obtain for bilayer graphene are 43.3 meV/atom and 50.6 meV/atom for AA and AB stacking, respectively. These values are in good agreement with the ones obtained from experimental works, namely,
 the exfoliation of graphite determines a binding energy between graphene layers of about 43 meV/atom,\cite{Girifalco} and that estimated for the separation of polyaromatic hydrocarbons is about 52 meV/atom. \cite{Zacharia}
As expected, we see in Fig. \ref{fig2} that the binding energy increases in absolute value with the nanoribbon width. Interestingly, the increase is not monotonous and for certain widths it shows an enhanced stability, see for instance the $N=4$ case. 

We find that the largest binding energy and most stable configuration for $N> 4$ is the AB$_\alpha$ in the AFM-{\it afm} configuration. Remarkably, for ultranarrow ribbons, $N=2,4$, the AA stacking with AFM-{\it afm} coupling is more favorable, albeit with a very close value of the BE to that of the  AB$_\alpha$ in the AFM-{\it afm} magnetic ordering. 

We now analyze in more detail the role of magnetic configurations on the stacking of ribbons.  The cases AB$_\alpha$ and AB$_\beta$ which have {\it fm} intralayer coupling are in the same energetic range, with the binding energy of AB$_\beta$ larger than AB$_\alpha$ for all widths. With the exception of the ultranarrow widths, all the AB$_\alpha$ cases, as well as the AB$_\beta$ with either FM or {\it fm} couplings, are rather close as to the binding energies.  This shows a relationship between stackings and magnetic configurations of the edges.

To elucidate the role of the magnetic interactions between edges, we have calculated the binding energies of ZGNRs on graphene. In such a case, because we are suppressing a ribbon, we are focusing on the edge-graphene interaction instead of the edge-edge interlayer interaction, and we only distinguish between AB and AA stackings.  We find that the binding energy is lower in Bernal stacking for all the studied widths, as it has
been
 previously found in bulk graphite, 
 where the Bernal stacking is more stable than the AA. This indicates that our finding on the greater stability and stronger binding energies for ultranarrow 4-ZGNR with AA stacking 
is related
to the
 edge-edge interlayer coupling.

\subsubsection{Structural and magnetic changes: quenching of the edge magnetic moments}

\begin{figure}[htbp] 
  \centering
    \includegraphics[width=8.5cm]{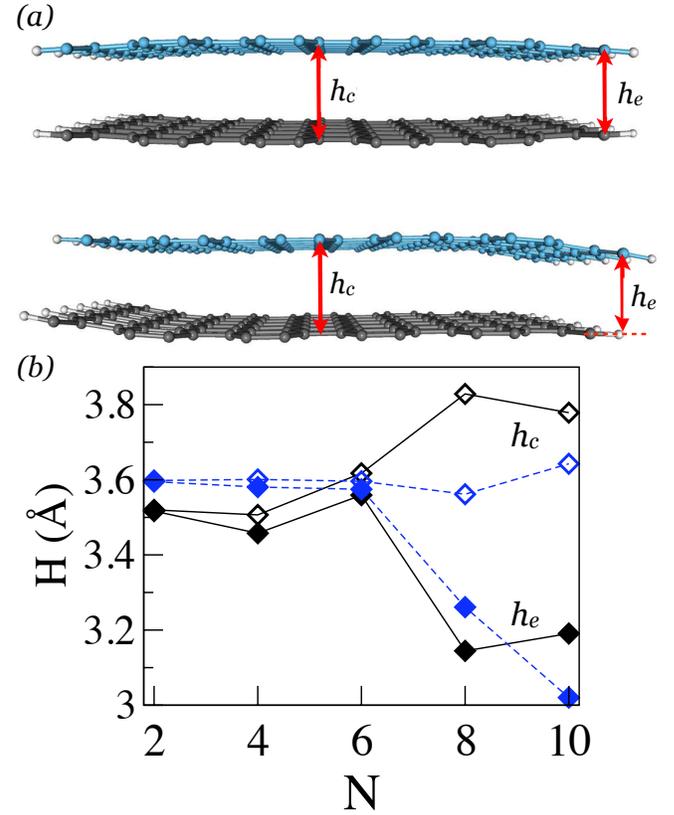}
  \caption{(Color online) (a) Structural distortions of the zigzag ribbon with $N=10$ for the most stable stackings AA and AB$_\alpha$ in the AFM-{\it afm} configurations after full relaxation.  The systems with higher binding energy (in absolute value) consist of non-planar, distorted, graphene ribbons.  (b) Interlayer distances at the center $h_c$ (empty diamonds) and edges h$_e$ (full diamonds) for {\it b}-ZGNRs with  AA (black) and AB$_\alpha$ [blue, gray] stackings in the AFM-{\it afm} configuration as a function of the ribbon width. 
}
  \label{fig3}
 \end{figure}

The differences in binding energies 
between
stacking orderings can be related to structural and magnetic changes in the bilayer ribbons.  In our fully relaxed simulations for {\it b}-ZGNR, all the AB$_\beta$ cases, and most of the AB$_\alpha$ cases, the layers remain flat.  Only the AA and AB$_\alpha$ stackings with AFM-{\it afm} coupling change their geometrical structure. Fig. \ref{fig3} (a) shows two examples for $N=10$.  Notice that the edges are bent inwards; in the AA case, the layers bend symmetrically, becoming convex at their center, 
and 
in the AB$_\alpha$ case, for which the ribbons are laterally displaced, the edges approach maintaining a flat central region.
The converged geometries of the ribbons are distorted, but still they have relevant symmetries, which are preserved within tolerance ($\approx 0.02 $ \AA):
the AA stacking with AFM-{\it afm} configuration shows a mirror symmetry and $C_2$ rotation with an axis parallel to the ribbons, while AB$_\alpha$ with AFM-{\it afm} shows only $C_2$ symmetry. 

\begin{figure}[htbp] 
  \centering
    \includegraphics[clip,width=6cm,angle=-90]{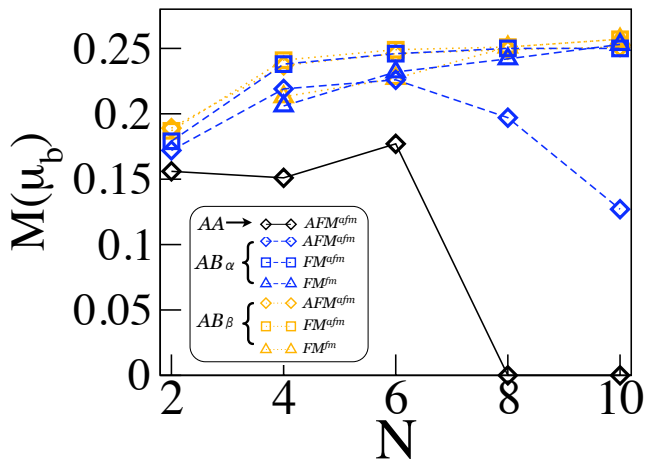}
  \caption{(Color online) Magnetization of the {\it b}-ZGNR as a function of the ribbon width $N$ after full relaxation.}
  \label{figmag}
 \end{figure}

To quantify these distortions,  in Fig. \ref{fig3} (b) we plot the distances between the two ribbons at their center, $h_c$, and at their edges, $h_e$, as a function of the bilayer width. The distances at the central part $h_c$ remain constant for the AB$_\alpha$ series, whereas they show larger changes for the AA stackings. For very small widths, up to $N=4$, the central distances $h_c$ for AA stackings are lower than for the AB cases. This is not what happens in bulk bilayer graphene, where 
the layer-to-layer distance is smaller for 
Bernal stacking than for AA.  
This
is an indication of the strong interaction in these ultranarrow ribbons with AA stacking.  For bilayer ribbons with $ N \geq 6$, the behavior is as expected, i.e., with central interlayer distances  $h_c$ smaller in the AB$_\beta$ ribbons than for the AA cases.  On the contrary, the distance between edges, $h_e$, is notably different from $h_c$ when N $>$ 6.  This deviation can be as large as 
0.6
\AA, indicating a strong edge-edge interaction.  Note that for the $AB_\alpha$ cases, these structural distortions are accompanied by a lateral sliding, but the values we find, e,g., 0.1 \AA\
  for $N=10$, are much smaller than those reported previously.  \cite{fazzio} These interlayer distances $h_c$ and  $h_e$ show that for ultranarrow ribbons, up to $N=6$, the bilayer ribbons 
behave rigidly,
becoming more flexible for larger widths.

 We now focus on the changes of magnetization at the edges $M$ due to the edge-edge interaction, as also shown in Fig. \ref{figmag}. It is defined as $M=N_{up}-N_{down}$, where $N_{up}$ ($N_{down}$) is the number of electrons with spin $up$ ($down$) per edge atom. 
The total magnetization by $95 \%$ corresponds to $p_z$ orbitals. We find that the magnetic moments are mainly located at the edges and decay exponentially when moving into the central part of the ribbon, in agreement with previous results.\cite{son_nature,mananes} Notice that the interlayer interaction between edges suppresses the 
site
magnetization: for the planar cases, the magnetization value is about 0.25 $\mu _b$. This is the case for all the magnetic configurations with AB$_\beta$ stacking,  and the AB$_\alpha$ ones with FM intralayer coupling, see Fig. \ref{figmag}.  However, for the AA and AB$_\alpha$ stackings with AFM-{\it afm} couplings, when the interlayer edge distance $h_e$ is reduced, a strong interaction appears between the $p_z$ orbitals at opposite edges, and the spin cloud evolves towards nonmagnetic configurations. In fact, the spin polarization for $ N\geq10$ is almost quenched.

\subsubsection{Implications of the magnetic quenching for calculations and experiments}

\begin{figure}[htbp] 
   \centering   
    \includegraphics[height=8.5cm]{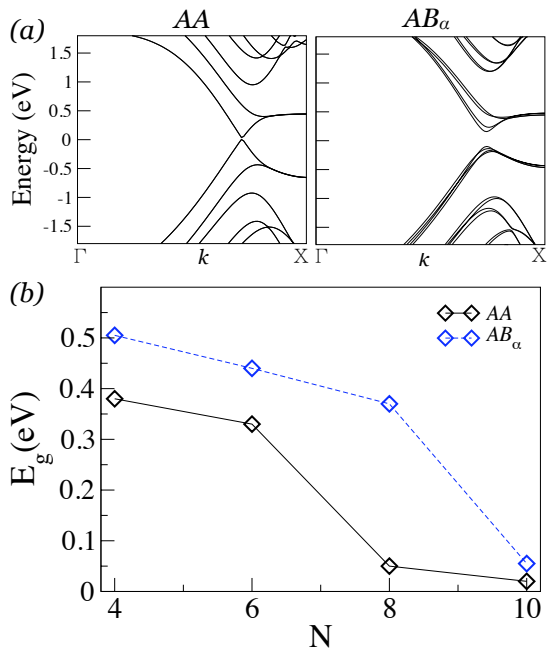}
   \caption{(Color online) (a) Band structure of a {\it b}-ZGNR  with $N=8$ in the most stable stackings, AA and AB$_\alpha$, and AFM-{\it afm} magnetic configuration. The Fermi energy is set to zero. Note the narrowing of the gap. (b) Gaps versus ribbon widths for the two most stable stackings and magnetic configurations, namely, AA and AB$_\alpha$ all with AFM-{\it afm}.}
   \label{fig4}
\end{figure}

{\it Narrowing of gaps at large widths}. 
These magnetic and structural changes are associated with other variations of the electronic properties of ribbons. The band structures of the {\it b}-ZGNRs of width $N=8$ for AA and AB$_\alpha$ stackings in the AFM-{\it afm} configuration are shown in Fig. \ref{fig4} (a). The gaps of the ribbons with AA stacking are smaller than those of the AB$_\alpha$ ones. When increasing the ribbon width $N$ the gap narrows as $\sim 1/N$.  \cite{banerjee} 
Note the sharp drop in the energy gap after $N=6$, related to the sudden decrease of the edge magnetization and the subsequent metallization of the bilayer nanoribbon, with $E_g < 0.05$ eV for $N=8$ in the AA stacking, barely visible in Fig. \ref{fig4} (a).\cite{son_nature,mananes}  
 It should be noted that the nanoribbon gaps may be underestimated in a GGA calculation.\cite{Heyd} The use of other functionals, such as the hybrid Heyd-Scuseria-Ernzerhof\cite{HSE} would correct this effect; in any case, the gap decrease with increasing width and the abrupt jumps associated with the magnetic changes will certainly hold in calculations employing other functionals, but with the quenching of magnetic moments taking place at larger widths. 

{\it Magnetoelastic switching}. Our results show a strong relationship between structure deformation and magnetic configuration. However, a question that remains is on the reversibility of structural changes with respect to the magnetic configuration, which can be relevant for experiments. To address it, as well as to corroborate the interplay between magnetism and structural changes, we have chosen the case of $N=10$ in the ground-state, i.e., stacking AB$_\alpha$ and AFM-{\it afm} configuration.
As it is shown in the previous Section, the ribbons in this structure are strongly curved. The application of a magnetic field perpendicular to the layer flips 
the magnetic moments at the edges from an AFM to an FM interlayer configuration; when we
flip the magnetic moments from AFM to FM interlayer coupling in the curved structure and we relax it, it converges to a planar geometry in the FM configuration. 
We consider this finding to be an indication of a 
change from curved to planar geometry driven by magnetic fields.
In principle, in a magneto-mechanical device based on these ribbons, one could control the edge deformation with magnetic fields, which in turn can produce other electronic changes such as gap narrowing.

 \begin{figure}[htbp] 
   \centering   
    \includegraphics[height=7.5cm,angle=-90]{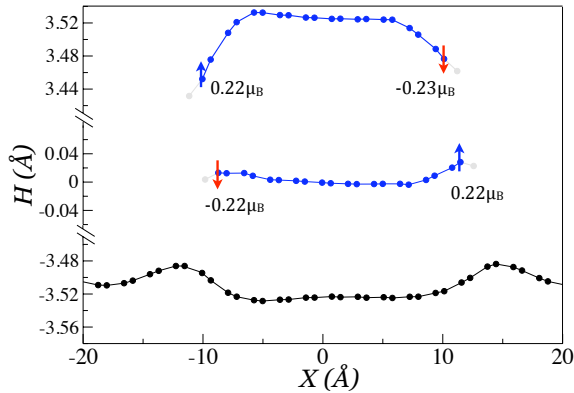}
   \caption{(Color online) Geometry of the {\it  b}-ZGNR with $N=10$ deposited on graphene. We have chosen a cut perpendicular to the graphene plane in order to highlight the edge deformations. The local magnetic moments on the edge atoms are also depicted. Note that the magnetic moments are almost as large as in a single ribbon.}
   \label{such}
\end{figure}

 {\it Effect of a substrate}.  We have studied the magnetic interaction of the edges in bilayer ribbons when deposited on a graphene substrate.  Figure \ref{such} shows the geometric and magnetic structure of the {\it b}-ZGNR with $N=10$ deposited on graphene in the AFM-{\it afm} magnetic configuration with AB$_\alpha$ stacking.
The interlayer distance in the center of the structure is overestimated, as it also occurs in bulk graphite\cite{LMKLL}; this is a consequence of the
vdW-DF2 functional used.
The top nanoribbon is deformed most, while the lower nanoribbon is nearly planar, due to the competing interaction between the graphene layer and the top ribbon. Due to this flat geometry,  the associated magnetization at the edges has higher values than the those obtained for bilayer nanoribbons, close to the ones of a single strip. As the structural deformation
of the sandwiched layer
is impeded
by the substrate interaction, the magnetic quenching is also precluded. Therefore, bilayer nanoribbons on substrates will remain magnetic for larger values of $N$ than
when
suspended. Bilayer nanoribbons with widths about $20$ nm can therefore act as a spintronic device, maybe not in the free standing geometry, but certainly on substrates.

\section{Summary} 

Bilayer zigzag graphene nanoribbons have been studied by first principles DFT calculations including a vdW-DF2 van der Waals functional. Four possible magnetic configurations have been explored for the three more symmetric stackings, 
the AA and two Bernal (AB$_\alpha$ and AB$_\beta$).

Our results show that the AA stacking is more favorable for ultranarrow ribbons in the AFM-{\em afm} configuration, competing in energy with the Bernal AB$_\alpha$  for larger {\it b}-ZGNRs.  The edge interaction bends their structure inwards for the AA and AB$_\alpha$ stackings with an intralayer and interlayer  antiferromagnetic configuration, but this bending is reduced for the smallest widths.  With increasing ribbon width,  the structural deformation at the edge is larger, leading to a reduction of the edge magnetic moments and the metallization of the {\it b}-ZGNRs.  A magnetic external field can modify the structural changes, flattening the ribbons.  We have also studied the effect of a graphene substrate.  In this case Bernal stacking is more favorable, and the bilayer ribbons maintain their magnetization for larger widths. This is due to the reduction of the structural deformation because of the graphene substrate.

\section{Acknowledgments}

This work has been partially supported by the Spanish DGES under Grants No. FIS2009-08744 and FIS2010-19609-C02-02, the Basque Departamento de Educaci\'on and the UPV/EHU (Grant No. IT-366-07), and the Nanoiker project (Grant No. IE11-304) under the ETORTEK program funded by the Basque Research Departament of Industry. L. C. and H. S. gratefully acknowledge helpful discussions with Juan Manuel Rodr\'{\i}guez Puerta and the hospitality of the Donostia International Physics Center. Calculations were partly done using the Camgrid high-throughput facility of the University of Cambridge. H. S. acknowledges the hospitality of the Department of Earth Sciences of the University of Cambridge.

\end{document}